\documentclass[APS,twocolumn,amsmath,amssymb]{revtex4} 

\begin{document}

\title{Propagation of ultrashort electromagnetic
wave packets in free space beyond the paraxial and
slowly-varying-envelope approximations : a time-domain approach}

\author{Charles Varin and Michel Pich\'e}
\affiliation{Centre d'optique, photonique et laser, Universit\'e
Laval, \\Qu\'ebec, G1K 7P4, Canada}%

\begin{abstract}A time-domain
approach is proposed for the propagation of ultrashort
electromagnetic wave packets beyond the paraxial and the
slowly-varying-envelope approximations. An analytical method based
on perturbation theory is used to solve the wave equation in free
space without resorting to Fourier transforms. An exact solution
is obtained in terms of successive temporal and spatial
derivatives of a monochromatic paraxial beam. The special case of
a radially polarized transverse magnetic wave packet is discussed.
\end{abstract}



\maketitle 

State-of-the-art mode-locked laser oscillators can generate
ultrafast pulses whose duration can be as short as a few optical
cycles~\cite{brabecRMP2000}. Such pulses -- with a finite
spatiotemporal extent and thus carrying a finite quantity of
energy -- can in principle be isolated, amplified, temporally
compressed, and spatially focused to reach extreme field
intensities within a volume that is of the order of the third
power of the laser wavelength ($\lambda^3$)~\cite{brabecRMP2000}.
In this situation the commonly used paraxial approximation and
slowly-varying-envelope approximation (SVEA) do not apply directly
and must be corrected.

In 1975, Lax \textit{et al.} proposed a method to express exact
solutions of Maxwell's equations in terms of an infinite series of
corrections applied to a paraxial electric
field~\cite{laxPRA1975}. So far, this approach has led numerous
studies dealing with non-paraxial optics. Unfortunately, as it is
also the case for the electromagnetic beam theory developed by
Davies~\cite{davisPRA1979}, the so-called ``method of Lax'' only
applies to monochromatic beams, or CW laser signals. With the
outstanding developments in femtosecond and attosecond
technologies, more appropriate tools are required to describe the
propagation of non-paraxial ultrashort light pulses, sometimes
called light bullets. Recently, Lu \textit{et
al.}~\cite{luJPA2003} have proposed a method to describe the
free-space propagation of vectorial non-paraxial ultrashort pulsed
beams in the frequency domain. In this letter, we present an
alternative approach to this problem entirely in the time domain,
i.e. without using Fourier transforms.

The spatiotemporal propagation of an electromagnetic wave packet
is governed by Maxwell's equations. Solving these equations in
free space proceeds in three steps~: 1. the transverse electric
field vector $\mathbf{E}_\bot$ is found from the wave equation, 2.
the longitudinal electric field $E_z$ is obtained from the
divergence equation ($\nabla\cdot\mathbf{E}=0$), and 3. the
magnetic flux vector $\mathbf{B}$ is calculated from the
differential form of Faraday's law
($\nabla\times\mathbf{E}=-\partial_t\mathbf{B}$) or the
differential form of Amp\`ere's law
($\nabla\times\mathbf{B}=c^{-2}\partial_t\mathbf{E}$), where $c$
is the speed of light \textit{in vacuo}, and
$\partial_{\,t}\equiv\partial/\partial t$. In the case of a
paraxial wave packet whose duration is much longer than the
optical period, this sequence of operations is performed with
ease. We now describe each of these three steps in detail beyond
the range of validity of the paraxial approximation and of the
SVEA.

In free space, the vectorial wave equation for the transverse
electric field vector $\mathbf{E}_\bot$ -- perpendicular to the
time-averaged Poynting vector -- is given by the following wave
equation~\cite{jackson}:
\begin{equation}\label{eq:wave_equation}
\nabla^2\mathbf{E}_\bot-\frac{1}{c^2}\,\partial_{\,t}^{\,2}
\,\mathbf{E}_\bot=0.
\end{equation}
The electric field vector is defined as follows in phasor notation
(see also Refs.~\onlinecite{brabecRMP2000} and
\onlinecite{brabecPRL1997})~:
\begin{eqnarray}\label{eq:E_phasor}
\mathbf{E}&=&\mathbf{E}_\bot+E_z\mathbf{a}_z\nonumber,\\
&=&\mathrm{Re}\left[\tilde{\mathbf{E}}\exp\left[
j(\omega_0t-k_0z)\right]\right]\nonumber,\\
&=&\mathrm{Re}\left[\left(\tilde{\mathbf{E}}_\bot+\tilde{E}_z\mathbf{a}_z
\right)\exp\left[j(\omega_0t-k_0z)\right]\right],
\end{eqnarray}
where $\mathbf{\tilde{E}}_\bot$ is the complex envelope of the
transverse electric field vector, $\tilde{E}_z$ is the complex
envelope of the longitudinal electric field, $\mathbf{a}_z$ is a
longitudinal unit vector oriented along the pulse propagation,
$j=\sqrt{-1}$ ; $\omega_0=k_0c$, $k_0=2\pi\lambda_0$, and
$\lambda_0$ are the central angular frequency, the central wave
number, and the central wavelength of the wave packet spectrum,
respectively. Combining Eqs.~(\ref{eq:wave_equation})
and~(\ref{eq:E_phasor}), one obtains the following equation for
the complex envelope of the field~:
\begin{equation}\label{eq:wave_equation_envelope}
\nabla_\bot^2\mathbf{\tilde{E}}_\bot-2jk_0\partial_{\,z}\mathbf{\tilde{E}}_\bot
-2j\frac{\omega_0}{c^2}\partial_{\,t}\mathbf{\tilde{E}}_\bot
+\partial_{\,z}^{\,2}\mathbf{\tilde{E}}_\bot
-\frac{1}{c^2}\partial_{\,t}^{\,2}\mathbf{\tilde{E}}_\bot=0,
\end{equation}
where $\nabla_\bot$ is a differential operator acting in the
transverse plane.

The use of the retarded time $t'=t-z/c$ instead of the (normal)
time $t$ is sometimes preferred when dealing with the propagation
of electromagnetic signals of finite spatiotemporal
extent~\cite{jackson}. More specifically, it separates the
transverse envelope (diffraction) from the temporal envelope
(pulse shape). After some manipulations, the result given at
Eq.~(\ref{eq:wave_equation_envelope}) can be expressed in terms of
the retarded variables $t'=t-z/c$ and $z'=z$ as follows~:
\begin{equation}\label{eq:wave_equation_envelope_retarded}
\nabla_\bot^2\mathbf{\tilde{E}}_\bot-2jk_0\partial_{\,z'}\mathbf{\tilde{E}}_\bot
+\partial_{\,z'}^{\,2}\mathbf{\tilde{E}}_\bot
-\frac{2}{c}\,\partial_{\,z'}\partial_{\,t'}\mathbf{\tilde{E}}_\bot=0.
\end{equation}
Alternatively, Eq.~(\ref{eq:wave_equation_envelope_retarded}) can
be written in a more compact form that reads~:
\begin{equation}\label{eq:wave_equation_envelope_retarded_perturbated}
\nabla_\bot^2\mathbf{\tilde{E}}_\bot-2jk_0\partial_{\,z'}\left[\,
1-\Theta\,\right]\mathbf{\tilde{E}}_\bot =0,
\end{equation}
with
$\Theta=j\left(\omega_0^{-1}\partial_{\,t'}-(2k_0)^{-1}\partial_{\,z'}\right)$.
For paraxial and quasi-monochromatic optical beams, the term
$\Theta\mathbf{\tilde{E}}_\bot$ is vanishingly small compared to
the complex field $\mathbf{\tilde{E}}_\bot$ itself, i.e.
$\left[\mathbf{\tilde{E}}_\bot-\Theta\mathbf{\tilde{E}}_\bot\right]
\simeq\mathbf{\tilde{E}}_\bot$. In that case,
Eq.~(\ref{eq:wave_equation_envelope_retarded_perturbated}) leads
to the \textit{paraxial wave equation}~\cite{siegman1986}. On the
other hand, when the transverse (or longitudinal) dimension of the
wave packet is of the order of the wavelength, the term
$\Theta\mathbf{\tilde{E}}_\bot$ can no longer be neglected.
However, the modulus of this contribution, i.e.
$|\Theta\mathbf{\tilde{E}}_\bot|$, remains small even for strongly
focused single-cycle light
pulses~\cite{{brabecPRL1997},{laxPRA1975},{porrasOP2001},{agrawalPRA1983}}.
As a consequence, we can consider
Eq.~(\ref{eq:wave_equation_envelope_retarded_perturbated}) to be a
\textit{perturbed paraxial wave equation}.

We observe that the partial differential operator $\Theta$ cannot
be reduced to a simpler operator, proportional only to spatial or
temporal derivatives, i.e. $\Theta\propto\partial_{\,z}^{\,n}$ or
$\Theta\propto\partial_{\,t}^{\,n}$. This fact clearly indicates
that spatial and temporal effects cannot be decoupled.
Consequently, a solution of
Eq.~(\ref{eq:wave_equation_envelope_retarded_perturbated}) is
found by expanding the transverse electric field vector
$\mathbf{\tilde{E}}_\bot$ as a power series of $\Theta$. In a
compact notation it reads~:
\begin{equation}\label{eq:E_trans_envelope_expansion}
\mathbf{\tilde{E}}_\bot=\sum_{n=0}^{\infty}\Theta^{n}
\mathbf{\tilde{\Psi}}_\bot^{(n)}.
\end{equation}
By equating terms with the same power of $\Theta$ in
Eq.~(\ref{eq:wave_equation_envelope_retarded_perturbated}), the
two following equations are then found~:
\begin{subequations}
\begin{eqnarray}\label{eq:wave_equation_paraxial}
\nabla_\bot^2\mathbf{\tilde{\Psi}}_\bot^{(0)}-2jk_0\partial_{\,z'}
\mathbf{\tilde{\Psi}}_\bot^{(0)}&=&0,\\\label{eq:wave_equation_correction_nth_term}
\nabla_\bot^2\mathbf{\tilde{\Psi}}_\bot^{(n)}
-2jk_0\partial_{\,z'}\mathbf{\tilde{\Psi}}_\bot^{(n)}+2jk_0\partial_{\,z'}
\mathbf{\tilde{\Psi}}_\bot^{(n-1)}&=&0,
\end{eqnarray}
\end{subequations}
solving the \textit{perturbed paraxial wave equation} in such a
way that the $n$th contribution is obtained recursively from the
order $(n-1)$, acting like a source term in the paraxial wave
equation ($\mathbf{\tilde{\Psi}}_\bot^{(n<0)}=0$).

A solution of the recursion relation given at
Eq.~(\ref{eq:wave_equation_correction_nth_term}) is written as
follows (see Porras\cite{porrasOP2001})~:
\begin{subequations}
\begin{eqnarray}\label{eq:Psi_n_solution_de_base}
\mathbf{\tilde{\Psi}}_\bot^{(n)}&=&\partial_{\,z'}^{\,n-1}\left(
\frac{z'^{\,n}}{n!}\,\partial_{\,z'}\mathbf{\tilde{\Psi}}_\bot^{(0)}\right),\\
\label{eq:Psi_n_solution_de_base_avec_somme}&=&\sum_{m=1}^{n}{n-1
\choose m-1}\frac{\,z'^m}{m!}
\,\partial_{\,z'}^{\,m}\mathbf{\tilde{\Psi}}_\bot^{(0)},
\end{eqnarray}
\end{subequations}
thus giving $\mathbf{\tilde{\Psi}}_\bot^{(n)}$ in terms of
successive derivatives of the zero-order field
$\mathbf{\tilde{\Psi}}_\bot^{(0)}$, a solution of the paraxial
wave equation. Recalling
Eq.~(\ref{eq:E_trans_envelope_expansion}), the envelope of the
transverse electric field vector reads~:
\begin{equation}\label{eq:E_trans_envelope_expansion_solution}
\mathbf{\tilde{E}}_\bot=\sum_{n=0}^{\infty} j^{\,n}
\left(\frac{1}{\omega_0}\partial_{\,t'}-\frac{1}{2k_0}
\partial_{\,z'}\right)^n\partial_{\,z'}^{\,n-1}\left(
\frac{z'^{n}}{n!}\,\partial_{\,z'}\mathbf{\tilde{\Psi}}_\bot^{(0)}\right).
\end{equation}

For any given transverse electric field vector $\mathbf{E}_\bot$,
the longitudinal electric field $E_z$ can be obtained from the
divergence equation. With the use of the complex notation defined
at Eq.~(\ref{eq:E_phasor}), $\nabla\cdot\mathbf{E}=0$ can be
formally inverted~\cite{{haselhoffPRE1994}} and expressed in terms
of the retarded variables $t'$ and $z'$ to yield~:
\begin{equation}\label{eq:Ez_complex}
\tilde{E}_z=\frac{-j\,}{k_0}\sum_{m=0}^{\infty}j^{\,m}
\left(\frac{1}{\omega_0}\partial_{\,t'}-\frac{1}{k_0}
\partial_{\,z'}\right)^m\left(\nabla_{\bot}\cdot\tilde{\mathbf{E}}_\bot\right).
\end{equation}

A general equation giving the magnetic flux vector $\mathbf{B}$ of
an arbitrary ultrashort wave packet can be deduced from Maxwell's
equations but cannot always be reduced to a simple expression, as
it is the case for monochromatic transverse electromagnetic (TEM)
waves (see Eq. (7.11) of Ref.~\onlinecite{jackson}). However, for
a given distribution and polarization of the transverse electric
field vector corresponds a spatiotemporal arrangement of the
fields that is unique and guarantees the stability of the wave
packet. As an example, let us consider the case of a TM$_{0l}$
wave packet whose transverse electric field is radially polarized,
i.e. $\mathbf{E}_\bot=E_r\mathbf{a}_r$ and $E_\theta=0$, where $r$
and $\theta$ are polar coordinates in the plane perpendicular the
propagation axis (the $z$-axis), and $l=1,2,3,\ldots$ The
intensity profile of this particular family of beams is
characterized by $l$ concentric and angularly symmetric
($\partial_\theta\mathbf{E}_\bot=0$) bright
rings~\cite{siegman1986}. From Amp\`ere's law
($\nabla\times\mathbf{B}=c^{-2}\partial_t\mathbf{E}$), we obtain
the two following equations ($B_r=0$ and $B_z=0$)~:
\begin{subequations}
\begin{eqnarray}\label{eq:ampere_law_TM01}
-\partial_{\,z} B_\theta&=&c^{-2}\partial_{\,t}E_r,\label{eq:ampere_law_TM01_Er}\\
r^{-1}\partial_{\,r}(r
B_\theta)&=&c^{-2}\partial_{\,t}E_z.\label{eq:ampere_law_TM01_Ez}
\end{eqnarray}
\end{subequations}
If we combine Eqs.~(\ref{eq:ampere_law_TM01_Er})
and~(\ref{eq:ampere_law_TM01_Ez}) together, we obtain the
divergence equation that we solved earlier at
Eq.~(\ref{eq:Ez_complex}). Thus, the solution of
Eq.~(\ref{eq:ampere_law_TM01_Er}) assures that
Eq.~(\ref{eq:ampere_law_TM01_Ez}) is also respected. If we express
the fields in complex notation (and then in terms of the retarded
variables), the formal inversion of
Eq.~(\ref{eq:ampere_law_TM01_Er}) yields~:
\begin{equation}\label{eq:B_theta_TM01_series}
\tilde{B}_\theta=\frac{1}{c}\,\sum_{p=0}^{\infty}j^{\,p}
\left(\frac{1}{\omega_0}\partial_{\,t'}-\frac{1}{k_0}
\partial_{\,z'}\right)^p
(1-j/\omega_0\partial_{\,t'})\,\,\tilde{E}_r.
\end{equation}

In the past, many investigations have been devoted to the study of
the paraxial wave equation and its solutions (see for example
Ref.~\onlinecite{eriksonPRE1994}). More specifically, we make
reference to the wide family of the Gauss-Hermite and
Gauss-Laguerre beam modes, considered to be a good representation
of the transverse electromagnetic modes of optical
resonators~\cite{{davisPRA1984},{siegman1986}}. According to the
formalism we propose, these solutions, or any solution obtained
under the paraxial beam optics theory, can be extended to the case
of tightly focused (non-paraxial) ultrashort (few-cycle) wave
packets using the general formulation we developed. We emphasize
the fact that we obtained all the spatiotemporal corrections in
the time-domain, alternatively to the Fourier transform
approach~\cite{luJPA2003}. However, the method described here has
the feature that it displays directly the corrections of the SVEA
on the four-dimensional structure of the wave packet, as needed to
correctly predict ultrafast interactions with matter. For
instance, dispersion and nonlinear optical effects could be
embedded in this formalism~\cite{brabecPRL1997}.

This letter should be considered as an extension  -- a vectorial
and non-paraxial generalization -- of a previous work published by
Porras~\cite{porrasOP2001}. In fact, it is possible to show that
Eq.~(\ref{eq:E_trans_envelope_expansion_solution}) simplifies to
the equation for paraxial ultrashort pulsed beams when the spatial
variations of the field due to diffraction are negligible compared
to the variations of the temporal envelope, i.e. when
$\Theta\simeq j\omega_0^{-1}\partial_{\,t'}$ (see Eq.~(8) of
Ref.~\onlinecite{porrasOP2001}). Besides, it has been demonstrated
that the perturbation method we used leads to the same results
that are obtained from integral methods~\cite{porrasOP2001}. We
thus observe that the fields given at
Eqs.~(\ref{eq:E_trans_envelope_expansion_solution})
and~(\ref{eq:Ez_complex}) are equivalent to the Eqs.~(25) and~(28)
of the paper of Lu~\textit{et al.}~\cite{luJPA2003}.

The irreducibility of the expansion parameter $\Theta$ indicates
that in free space, the spatial and temporal envelopes of a
non-paraxial ultrashort wave packet are nonseparable. This
coupling ensures the wave packet's stability under tight focusing
and reveals itself as a reorganization of the structure of the
wave packet, noticeable as modifications of the carrier wave and
the spatiotemporal envelope~\cite{porrasOP2001}.

For a gaussian beam (with a gaussian pulse shape), the spatial and
temporal contributions, i.e.
$(2k_0)^{-n}\partial_{\,z'}^{\,n}\tilde{\mathbf{E}}_\bot$ and
$\omega_0^{-n}\partial_{\,t'}^{\,n}\tilde{\mathbf{E}}_\bot$, are
respectively proportional to $(2k_0 z_R)^{-n}=(k_0 r_0)^{-2n}$ and
$(\omega_0\Delta t)^{-n}$, where $z_R=k_0r_0^2/2$ is the Rayleigh
distance, $r_0$ is the wave packet transverse extent at focus
(beam waist), and $\Delta t$ is the pulse duration. The expansion
of $\Theta^n$ yields also crossed-terms with various powers of
$k_0 r_0$ and $\omega_0\Delta t$. However, for a single-cycle
pulse ($\Delta t=2\pi/\omega_0$) focused down to one wavelength
($r_0=2\pi/k_0$), the most important contribution comes from the
temporal envelope and decreases as $(2\pi)^{-n}$. Consequently, it
is observed that a truncated series ($n=0,1,2$) reproduces the
exact field with a good accuracy (see Porras~\cite{porrasOP2001}).
Nonetheless, for slightly focused and long pulses ($r_0>>2\pi/k_0$
and $\Delta t>>2\pi/\omega_0$), the formalism reduces to the
expressions for paraxial and monochromatic
beams~\cite{{laxPRA1975},{eriksonPRE1994}}.

The full vectorial treatment we have presented here also leads to
an exact evaluation of the associated longitudinal electric and
magnetic fields. These field components are not usually dealt with
in optics ; however, they must be taken into consideration for the
study of relativistic effects in laser-matter interactions and for
the investigation of electron acceleration in intense laser
fields.

\section*{Acknowledgments}
M. Pich\'e and C. Varin thank Les fonds de recherche sur la nature
et les technologies (Qu\'ebec), the Natural Sciences and
Engineering Research Council (Canada), and the Canadian Institute
for Photonic Innovations for their financial support. C. Varin
also thank Miguel A. Porras for helpful discussions.

\end{document}